\documentclass[11pt,twoside]{article}
\usepackage{./asp2014}

\aspSuppressVolSlug
\resetcounters

\begin{document}%%%%%%%%%%%%%%%%%%%%%%%%%%%%%%%%%%%%%%%%%%%%%%%%%%%%%%%%%%%%%%%%%%%%%%%%%%%%%%%%%%%%%%%%%%%%%%%%%%%%%%%%%%%%%%%%%%%%%%%%%%%%%%%%%%%%
%%%%%%%%%%%%%%%%%%%%%%%%%%%%%%%%%%%%%%%%%%%%%%%%%%%%%%%%%%%%%%%%%%%%%%%%%%%%%%%%%%%%%%%%%%%%%%%%%%%%%%%%%%%%%%%%%%%%%%%%%%%%%%%%%%%%%%%%%%%%%%%%%%%%%

\title{Observational studies of transiting extrasolar planets}
\author{John~Southworth
\affil{Astrophysics Group, Keele University, Staffordshire, ST5 5BG, UK}\email{astro.js@keele.ac.uk}}

\paperauthor{John~Southworth}{astro.js@keele.ac.uk}{}{Keele University}{Astrophysics Group}{Newcastle-under-Lyme}{Staffordshire}{ST5 5BG}{UK}

\begin{abstract}
The study of transiting extrasolar planets is only 15 years old, but has matured into a rich area of research. I review the observational aspects of this work, concentrating on the discovery of transits, the characterisation of planets from photometry and spectroscopy, the {\it Homogeneous Studies} project, starspots, orbital obliquities, and the atmospheric properties of the known planets. I begin with historical context and conclude with a glance to a future of TESS, CHEOPS, {\it Gaia} and PLATO.
\end{abstract}

%%%%%%%%%%%%%%%%%%%%%%%%%%%%%%%%%%%%%%%%%%%%%%%%%%%%%%%%%%%%%%%%%%%%%%%%%%%%%%%%%%%%%%%%%%%%%%%%%%%%%%%%%%%%%%%%%%%%%%%%%%%%%%%%%%%%%%%%%%%%%%%%%%%%%

\section{History and context}

The first widely accepted detection of an extrasolar planet orbiting a normal star was made by \citet{MayorQueloz95nat}, using high-precision radial velocity (RV) measurements. They found an object with a minimum mass of $M_{\rm b} \sin i = 0.47 \pm 0.02$\,M$_{\rm Jup}$ orbiting the solar-like star 51\,Peg every 4.2\,days. Earlier discoveries had been made, but were either treated with caution, had a significantly larger mass, or were orbiting pulsars (see \citealt{WrightGaudi13book} for an historical account). The second 51\,Peg-type planetary system followed quickly afterwards \citep{MarcyButler96apj} and by the start of the year 2000 a total of 25 planets had been detected, all by the RV method. Whilst valuable discoveries, only their minimum mass, orbital period, eccentricity and semimajor axis could be measured; their radius and thus density were unattainable.

One of the early RV planets was HD\,209458, and in late 1999 it was found to transit its host star \citep{Henry+00apj,Charbonneau+00apj}. Transiting extrasolar planets (TEPs) are intrinsically more useful because the depth of the transit depends on the planetary radius, ultimately allowing measurement of its density, surface gravity and true mass. The second known TEP was unveiled three years later and in a very different way, by RV follow-up of a star showing transits \citep{Konacki+03nat}.

\begin{figure}[t] \includegraphics[width=\textwidth,angle=0]{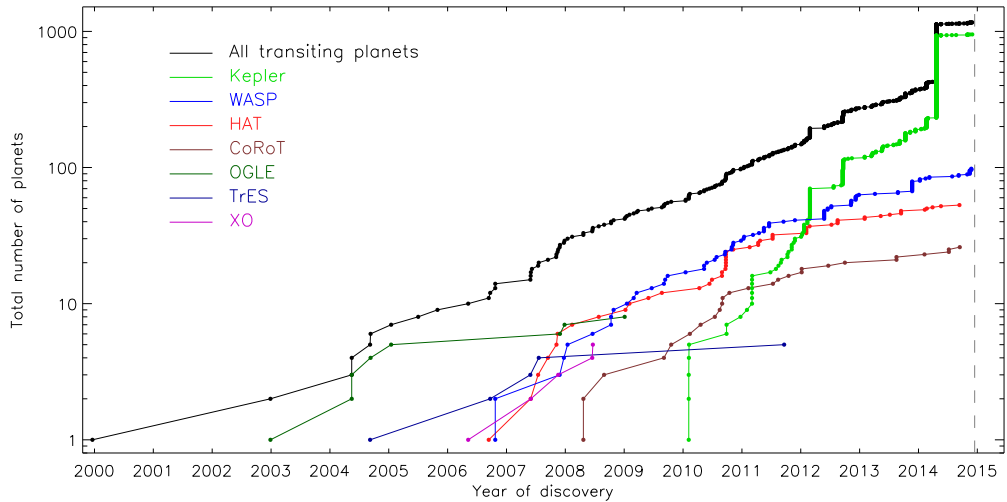} \\[-15pt]
\caption{\label{fig:discrate} The discovery rate of the known TEPs, illustrating the roughly
exponential growth. The coloured lines show the breakdown for each consortium.} \end{figure}

Whilst the initial rate of discovery of exoplanets was slow, it has shown exponential growth and now exceeds 1800 objects of which over 1150 are transiting\footnote{Data from TEPCat \citep{Me11mn} at: {\tt http://www.astro.keele.ac.uk/jkt/tepcat/}}. Fig.\,\ref{fig:discrate} shows the discovery rate of the known TEPs and breaks this down into the contributions from different consortia. The roughly exponential discovery rate gives a constant slope in this logarithmic plot, with the exception of the 851 planets in 340 multiple systems which were statistically validated by \citet{Rowe+14apj} in early 2014. The greatest number of discoveries have come from the {\it Kepler} satellite \citep{Borucki+10sci}, whose large aperture and space-based location yielded data of extremely high precision, duty cycle, and time coverage. The second most productive consortium is SuperWASP \citep{Pollacco+06pasp}, followed by HAT \citep{Bakos+02pasp}; these groups rely on small ground-based robotic telescopes equipped with telephoto lenses.

\begin{figure}[t] \includegraphics[width=\textwidth,angle=0]{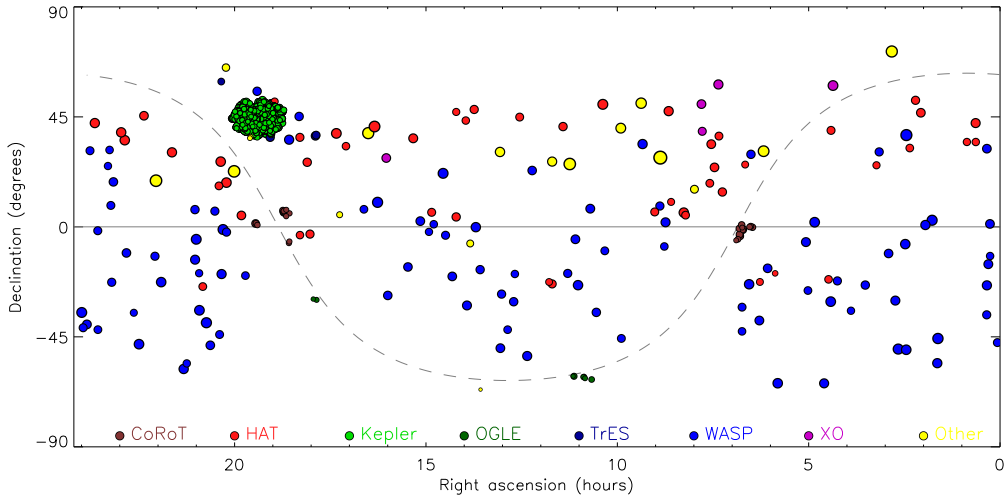} \\[-15pt]
\caption{\label{fig:skypos} The sky positions of the known TEPs, colour-coded according
to the discovery consortium (key along the base of the plot). The celestial equator is
shown with a grey solid line and the Galactic plane with a grey dashed line. Symbol size
depends on $V$-band apparent magnitude, with bright stars having larger point sizes.} \end{figure}

Fig.\,\ref{fig:skypos} shows the sky positions of the known TEPs, again colour-coded according to discovery consortium. The stand-out feature is the agglomeration of {\it Kepler} discoveries (green points at RA $=$ 19--20\,h and Dec = 40--50$^\circ$). The smaller brown groupings near the two intersections of the celestial equator and Galactic plane are due to the CoRoT satellite \citep{Moutou+13icar}, and the spread of blue in the Southern hemisphere come from the SuperWASP-South installation in South Africa.

\begin{figure}[t] \includegraphics[width=\textwidth,angle=0]{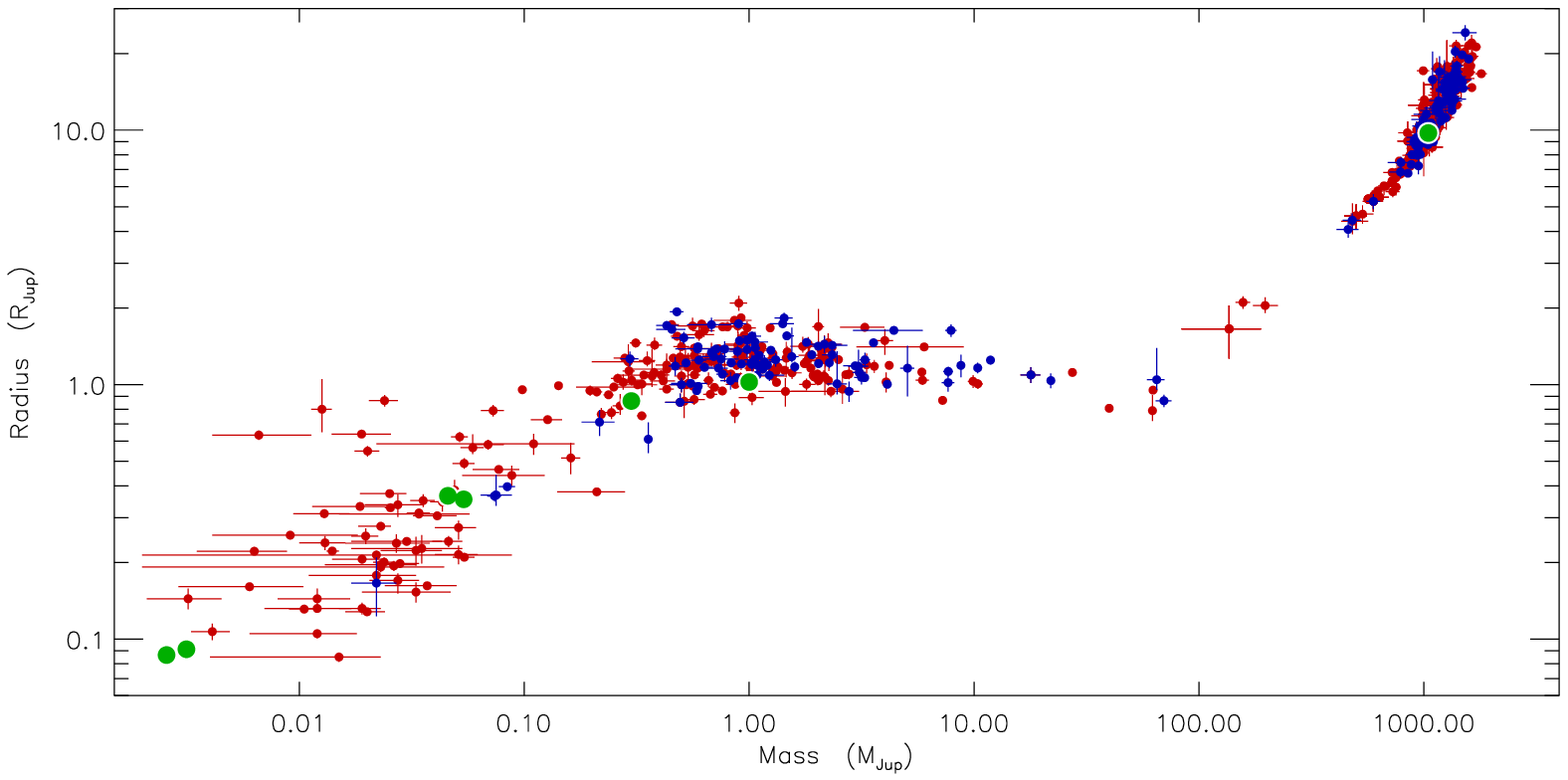} \\[-15pt]
\caption{\label{fig:mr} The mass-radius diagram for the known TEPs and their host stars.
Objects studied in the HSTEP project (Sect.\,\ref{sec:hstep}) are shown in blue, and other
objects are shown in red. The Solar-system bodies are indicated by green circles.} \end{figure}

Fig.\,\ref{fig:mr} shows the masses and radii of the known TEPs (main part of the diagram) and their host stars (dense assembly of points at the top-right). The fractional scatter in the properties of the planets is much more than that in the properties of their well-behaved FGK dwarf hosts, an indicator of the complexity of the physical effects which affect giant planets. The huge scatter in the properties of low-mass planets is due partly to the difficulty in characterising these small and low-mass objects, and partly to their extreme and poorly understood diversity \citep[e.g.][]{Masuda14apj}.

%%%%%%%%%%%%%%%%%%%%%%%%%%%%%%%%%%%%%%%%%%%%%%%%%%%%%%%%%%%%%%%%%%%%%%%%%%%%%%%%%%%%%%%%%%%%%%%%%%%%%%%%%%%%%%%%%%%%%%%%%%%%%%%%%%%%%%%%%%%%%%%%%%%%%

\section{Discovering and characterising transiting extrasolar planets}

Early work on the identification of TEPs concentrated mostly on the `hot Jupiters', which I consider to be gaseous planets of mass greater than 0.3\,M$_{\rm Jup}$ and orbital period less than 10\,d. These are the most easily identifiable planets because their relatively large radii lead to deep transits, and their masses and short orbital periods cause a comparatively large reflex velocity in the host stars. Eight TEPs were first identified using RV measurements and subsequently found to transit, including the two most-studied examples [HD\,209458 and HD\,189733 \citep{Bouchy+05aa}].

Planet detection via RV measurements is inherently expensive, requiring large telescopes and ultra-stable spectrographs, which are capable of observing only one target at once. As only a small fraction of stars host TEPs, this approach is an inefficient method of detection. The great majority of TEPs have therefore been found from large-scale photometric surveys, such as OGLE \citep{Udalski+02aca}, {\it Kepler}, WASP and HAT, which have the advantage of monitoring thousands of stars simultaneously. A major disadvantage of finding TEPs from photometric surveys is that not all transit events are due to planets. False positives can be caused by low-mass stars (late-M dwarfs have radii close to that of Jupiter; recall Fig.\,\ref{fig:mr}), faint eclipsing binaries whose light contaminates that of the target star, and instrumental effects. Planet candidates therefore have to be studied in detail to confirm their planetary nature.

{\it Kepler}'s space location and relatively high spatial resolution result in it having a low rate of false positives (see \citealt{MortonJohnson11apj} but also \citealt{Santerne+12aa} and \citealt{Coughlin+14aj}). For the CoRoT satellite, which has an inferior spatial resolution, $73 \pm7$\% of candidates are false positives and only 6\% are confirmed planets \citep{Moutou+13icar}, with the remainder being unsolved. The estimated false-positive rate for WASP-South is representative of a typical ground-based survey: roughly 1 in 14 candidates turns out to be of planetary mass \citep{Hellier+11conf}.

%%%%%%%%%%%%%%%%%%%%%%%%%%%%%%%%%%%%%%%%%%%%%%%%%%%%%%%%%%%%%%%%%%%%%%%%%%%%%%%%%%%%%%%%%%%%%%%%%%%%%%%%%%%%%%%%%%%%%%%%%%%%%%%%%%%%%%%%%%%%%%%%%%%%%

\subsection{Spectroscopic radial velocity measurements}

Once a transit event has been found, the planetary nature of the transiting object needs to be proved by measuring its mass. This can be done by obtaining multiple RV measurements using one of the current generation of high-resolution spectrographs such as Keck/HIRES, CORALIE or HARPS \citep[see][for a recent review]{Pepe+14nat}. The extremely high RV quality of which these instruments are capable allows the orbital motion of the host star to be measured. With some knowledge of the mass of the star, its orbital velocity amplitude ($K_{\rm A}$) indicates the mass of the transiting object\footnote{Subscripted letters `A' and `b' indicate properties of the host star and planet, respectively.}. The RVs also yield the planet's orbital eccentricity ($e$) and argument of periastron ($\omega$).

A bonus feature of the high-resolution spectra is that they can be used to determine the atmospheric parameters of the host star: its effective temperature ($T_{\rm eff}$), surface gravity ($\log g$) and metallicity ([M/H] or [Fe/H]). This process is typically achived by comparing the observed spectra to synthetic spectra either directly or via the measured equivalent widths of spectral lines \citep[e.g.][]{Torres+12apj}. These quantities, especially $T_{\rm eff}$, are vital for determining the mass of the star and thus the mass of the planet.

An alternative approach to RV measurements has been pursued for most of the {\it Kepler} planet candidates, necessitated by the faintness of most of these objects which makes high-resolution spectroscopy prohibitively expensive (often completely impossible) with current facilities. A large number of {\it Kepler} candidates have been `validated' by demonstrating the low probability of them being a false positive, instead of proving their planetary nature with a mass determination. The {\it Kepler} candidates are well suited to this approach because they are relatively small (too small to be a low-mass star) and very unlikely to be a result of contamination by a third object. The contamination can be investigated by high-resolution imaging and checking for apparent shifts in the position of the star during transit, effectively shrinking the sky area where contaminating objects can plausibly be located to a very small -- and therefore unlikely -- solid angle.

%%%%%%%%%%%%%%%%%%%%%%%%%%%%%%%%%%%%%%%%%%%%%%%%%%%%%%%%%%%%%%%%%%%%%%%%%%%%%%%%%%%%%%%%%%%%%%%%%%%%%%%%%%%%%%%%%%%%%%%%%%%%%%%%%%%%%%%%%%%%%%%%%%%%%

\subsection{Follow-up light curves} \label{sec:lc}

\begin{figure}[t] \includegraphics[width=\textwidth,angle=0]{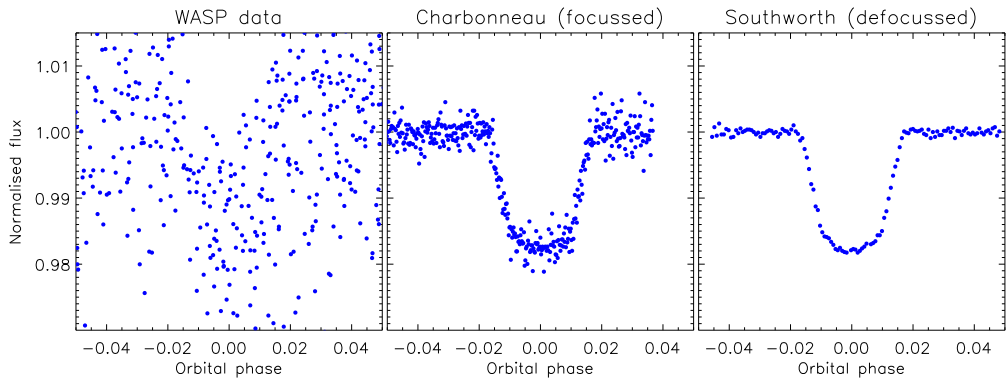} \\[-15pt]
\caption{\label{fig:wasp2} Example light curves of WASP-2. Left: SuperWASP data which led to its discovery
\citep{Cameron+07mn}. Centre: follow-up light curve from a 1.2\,m telescope operated in focus \citep{Charbonneau+07apj}.
Right: follow-up light curve from a defocussed 1.5\,m telescope \citep{Me+10mn}.} \end{figure}

Once a transiting object has been identified and proven to be of planetary origin via RV measurements, the next step is to obtain a high-quality light curve. The shape of the transit is a crucial piece of information for deducing the physical properties of the system, but discovery light curves from ground-based surveys are typically very scattered (see Fig.\,\ref{fig:wasp2}). A method of obtaining high-precision photometry which is now widely used is that of {\it telescope defocussing} \citep[e.g.][]{Alonso+08aa,Me+09mn}, whereby the point spread function (PSF) is broadened to cover hundreds or even thousands of pixels. There are two main advantages of this method. Firstly, flat-fielding noise is averaged down by the square-root of the number of pixels, i.e.\ several orders of magnitude. Secondly, longer exposure times are possible without saturating individual pixels, so less time is lost to reading out the CCD and more time is available to observe, thus decreasing the photon and scintillation noise.

\begin{figure}[t]
\includegraphics[width=0.49\textwidth,height=0.4\textwidth,angle=0]{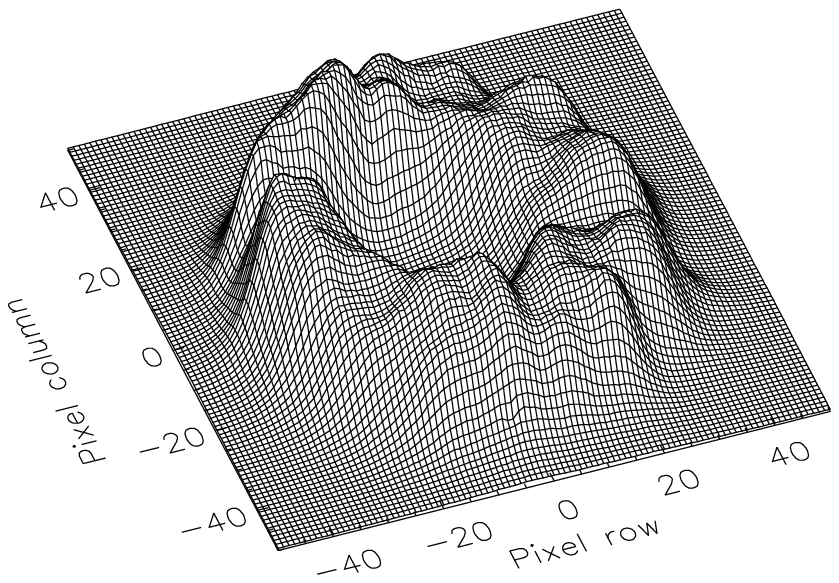}
\includegraphics[width=0.49\textwidth,height=0.4\textwidth,angle=0]{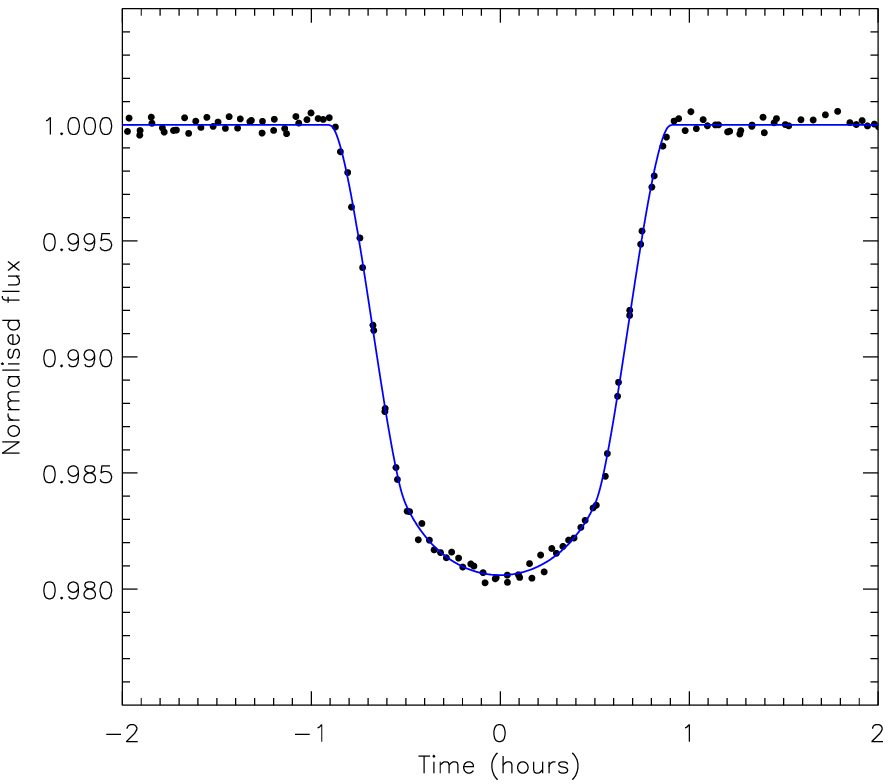} \\[-15pt]
\caption{\label{fig:wasp50} Left: example defocussed PSF of WASP-50 taken using NTT/EFOSC2.
Right: resulting light curve with a scatter of 0.24\,mmag \citep{TregloanSouthworth13mn}.
The line shows the best fit calculated using the {\sc jktebop} code.} \end{figure}

As an example of {\it telescope defocussing}, Fig.\,\ref{fig:wasp2} shows three light curves of the transit of WASP-2. The first panel shows the data used to detect the transit -- this was obtained using the SuperWASP-North installation which consists of 200\,mm telephoto lenses with a plate scale of 14$^{\prime\prime}$\,px$^{-1}$. The second panel shows an example of a follow-up light curve from a 1.2\,m telescope operated in focus \citep{Charbonneau+07apj}, reaching a very creditable scatter of 1.9\,mmag per point. The third panel displays a light curve obtained with a defocussed 1.5\,m telescope \citep{Me+10mn}, which achieves a scatter of only 0.46\,mmag per point. Fig.\,\ref{fig:wasp50} shows an example PSF and the resulting light curve of a transit of WASP-50 obtained with NTT/EFOSC2.

Once the shape of the transit has been observed, several important pieces of information can be extracted from it. Firstly, the depth of the transit is a strong indicator of the ratio of the radius of the planet to that of the star (a quantity called $k$), as the flux deficit indicates what fraction of the stellar surface is blocked by the dark planet. Secondly, the duration of the transit indicates how long it took the planet to pass in front of the star. This is closely related to the size of the star: the actual quantity measured is the fractional radius $r_{\rm A} = \frac{R_{\rm A}}{a}$ where $R_{\rm A}$ is the true radius of the stars and $a$ is the orbital semimajor axis. This quantity is often inverted and labelled $\frac{a}{R_\star}$. Thirdly, the duration of the partial phases of the transit (when only part of the planet is in front of the star) is a gauge for which part of the stellar disc the planet transits, i.e.\ the orbital inclination of the system ($i$). The orbital inclination is related to the impact parameter ($b$) by:
$$ b = \frac{1-e^2}{1 \pm e\sin\omega} \frac{\cos i}{r_{\rm A}} $$
where the $\pm$ is `$+$' for the transit and `$-$' for the occultation (secondary eclipse).

An important attribute of the fractional radius of the star is that it is very closely related to the stellar density, $\rho_{\rm A}$ \citep{SeagerMallen03apj}. From Kepler's third law and the definition of density we can derive the relation:
$$ r_{\rm A} = \frac{R_{\rm A}^{\,3}}{a^3} = \frac{3\pi}{GP^2} \frac{1}{\rho_{\rm A}} \left(\frac{M_{\rm A}}{M_{\rm A}+M_{\rm b}}\right) $$
where $G$ is the Newtonian gravitational constant, $P$ is the orbital period, and $M_{\rm A}$ and $M_{\rm b}$ are the masses of the star and planet. As $M_{\rm b} \ll M_{\rm A}$, the quantity in brackets can be ignored. An alternative formulation well suited to light curve analysis is:
$$ \rho_{\rm A} +  \left(\frac{R_{\rm A}}{R_{\rm b}}\right)^3 \rho_{\rm b} = \frac{3\pi}{GP^2} \left(\frac{a}{R_{\rm A}}\right)^{3}
\qquad \Rightarrow \qquad
\rho_{\rm A} +  k^3 \rho_{\rm b} = \frac{3\pi}{GP^2r_{\rm A}^{\,3}} $$
where $k$ is usually small, so $k^3$ is negligible,and the $k^3 \rho_{\rm b}$ term can be ignored.

The photometric parameters ($r_{\rm A}$, $k$ and $i$) can be obtained by fitting transit light curves with a simple geometric model such as the {\sc jktebop} program \citep{Me08mn,Me13aa} or the {\sc occultsmall} subroutine \citep{MandelAgol02apj}. The orbital ephemeris (period $P$ and reference time of mid-transit $T_0$) is easily obtained in the same way. There do, however, exist several complications.

{\bf Limb darkening} is one nuisance parameter which must be included in the model, and theoretically-derived coefficients are available for several approximation `laws' \citep[e.g.][]{ClaretBloemen11aa}. The use of theoretical coefficients is generally fine for data of ground-based but not of space-based quality \citep{Me08mn}.

{\bf Orbital eccentricity} affects the transit durations, because the orbital speed of the planet is no longer constant. It is essentially impossible to fit for this effect using only transit light curves \citep{Kipping08mn}. One must use the information provided by the RVs of the host star, either directly or by applying constraints to the light curve fit.

{\bf Cadence.} Some space-based light curves have a poor time sampling; most egregiously the {\it Kepler} long-cadence data with effective interagration times of 1765\,s \citep{Jenkins+10apj}. In these cases one must integrate the model to match the nature of the data or suffer potentially large errors in the results. \citet{Me11mn} showed that ignoring this problem gave photometric parameters wrong by 30\% for a typical case.

{\bf Contaminating light.} Faint stars close to TEP systems may contaminate light curves, causing the transit to be diluted and the planetary radius to be underestimated \citep{Daemgen+09aa}. This cannot be fitted for directly in the transit light curve \citep{Me10mn} as it is completely correlated with other parameters. But if faint stars can be detected using high-resolution imaging \citep[e.g.][]{Me+10mn,Lillobox++14aa}, their light can be accounted for in the model fit.

\begin{figure}[t]
\includegraphics[width=\textwidth,angle=0]{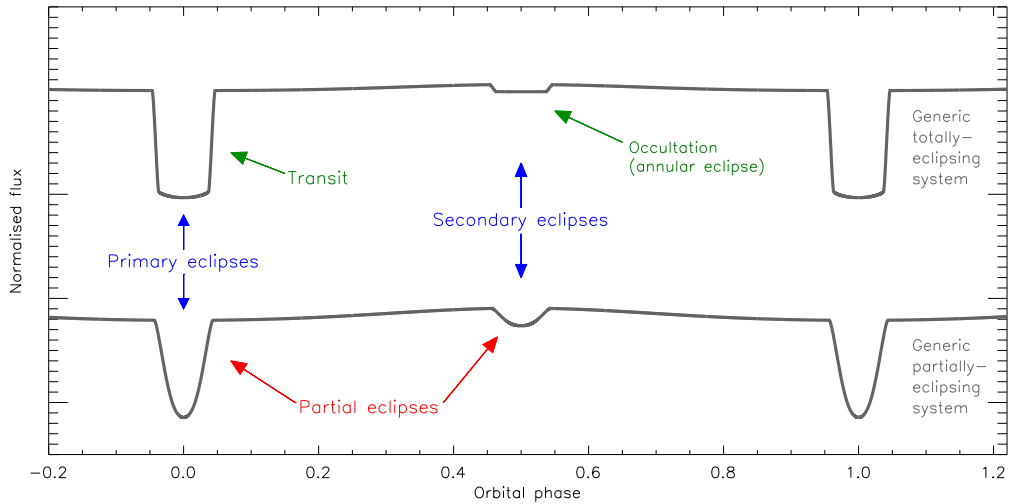} \\[-15pt]
\caption{\label{fig:term} Typical light curves of a totally-eclipsing and a partially-eclipsing
planetary system with the correct names of the eclipses indicated.} \end{figure}

I finish this section by discussing the terminology for the different types of eclipses seen in TEP systems. The correct terminology (see Fig.\,\ref{fig:term}) has been established for many years for for eclipsing binary systems \citep[e.g.][]{Hilditch01book}, and of course for solar and lunar eclipses. A `transit' is when a smaller object (e.g.\ planet) passes completely in front of a larger object (e.g.\ star). An `occultation' is when the planet passes behind the star. `Partial eclipses' can occur when part of one object never eclipses or is eclipsed by the other object. Eclipsing systems can have only one transit per orbit, so references to a `primary transit', `anti-transit' or `secondary transit' are incorrect.

%%%%%%%%%%%%%%%%%%%%%%%%%%%%%%%%%%%%%%%%%%%%%%%%%%%%%%%%%%%%%%%%%%%%%%%%%%%%%%%%%%%%%%%%%%%%%%%%%%%%%%%%%%%%%%%%%%%%%%%%%%%%%%%%%%%%%%%%%%%%%%%%%%%%%

\subsection{Determining the physical properties of transiting planets}

Fitting the RVs of the host star gives the parameters of the spectroscopic orbit: $K_{\rm A}$, $e$ and $\omega$. The combination terms $e\cos\omega$ and $e\sin\omega$ are often used instead of $e$ and $\omega$ themselves because they are less strongly correlated and not biased to higher values of $e$. Fitting the transit light curve gives the photometric parameters $r_{\rm A}$, $k$ and $i$. We also have extra information from the spectra of the host star: its $T_{\rm eff}$, $\log g$ and [M/H].

This situation is essentially that of an eclipsing binary system where only one star is seen in the spectra. The lack of RVs for the secondary component means $K_{\rm b}$ cannot be measured, so we are one piece of information short of being able to determine the physical properties of the system. Thankfully, an {\it additional constraint} can been obtained using the spectroscopic properties of the host star and either empirical calibrations of stellar properties or theoretical stellar evolutionary models. An elegant way to do this is to guess a value of $K_{\rm b}$ and use the other quantities ($P$, $K_{\rm A}$, $e$, $r_{\rm A}$, $k$ and $i$) to determine the mass and radius of the star and planet using standard formulae \citep[e.g.][]{Hilditch01book}. $M_{\rm A}$, $R_{\rm A}$ and $T_{\rm eff}$ can then be checked for consistency with the {\it additional constraint}, and $K_{\rm b}$ iteratively adjusted to maximise this consistency.

Most studies of TEPs have used theoretical stellar models to provide the {\it additional constraint}, in which case the advantage of conceptual simplicity is offset by the fact that it is not trivial to interpolate to arbitrary values within a tabulated grid of theoretical predictions. The reliance on stellar theory is worrying, as it is difficult to assess the effect of this on the results. One option is to try multiple sets of models and see how well they agree: \citet{Me10mn} found a {\em minimum} scatter of 1\% for $M_{\rm A}$, 0.6\% for $M_{\rm b}$ and less for other quantities. However, this only provides a lower limit on the true uncertainties because different sets of theoretical models have many areas of commonality such as computational approach, opacities and parameterisation of mixing.

An alternative to stellar theory is to construct (semi-)empirical calibrations of stellar properties based on the values measured for detached eclipsing binary systems (dEBs). This approach has its own advantange and disadvantage: the continuous nature of the calibrations means interpolation is not required, but it is not clear if the properties of low-mass stars are well-represented by dEBs \citep{Torres13an}. Calibrations were first used by \citet{Me09mn}, based on a simple mass-radius relation for late-type dwarfs. The problem with this approach is that the neglect of stellar evolution meant the results were not very reliable. A better approach was proposed by \citet{Torres++10aarv}, who calculated calibrations for stellar mass and radius as a function of $T_{\rm eff}$, $\log g$ and [Fe/H]. \citet{Enoch+10aa} further improved this approach by using $\rho_{\rm A}$ instead of $\log g$, motivated by the fact that $\rho_{\rm A}$ is directly obtained from transit light curves whereas $\log g$ can be inferred to only a lower precision by spectral analysis. Finally, \citet{Me11mn} followed the approach of \citet{Enoch+10aa} but based it on many more objects (180 versus 38 stars, sourced from the DEBCat\footnote{DEBCat \citep{Me14debcat} can be found at: {\tt http://www.astro.keele.ac.uk/jkt/debcat/}} catalogue of measured physical properties of well-studied dEBs).

Several quantities can be measured without requiring the {\it additional constraint}. The stellar density, $\rho_{\rm A}$, was already discussed in Sect.\,\ref{sec:lc}. The surface gravity of the planet can be obtained using only measured quantities \citep{Me++07mn}:
$$ g_{\rm b} = \frac{2\pi}{P} \,\frac{\sqrt{1-e^2}\, K_{\rm A}}{r_{\rm b}^{\ 2} \sin i} $$
where $r_{\rm b} = \frac{R_{\rm b}}{a}$ is the fractional radius of the planet.

\noindent The planetary equilibrium temperature is also independent of the scale of the system:
$$ T_{\rm eq} = T_{\rm eff} \sqrt{\frac{R_{\rm A}}{a}} \,\, \bigg[f(1-A_{\rm B})\bigg]^{1/4} = T_{\rm eff} \sqrt{r_{\rm A}} \,\, \Big[f(1-A_{\rm B})\Big]^{1/4} $$
where $A_{\rm B}$ is the Bond albedo and $f$ is the heat redistribution parameter \citep[e.g.][]{SheetsDeming14apj}. A common approach is to assume $f(1-A_{\rm B}) = 1$ in which case the equation becomes very simple: $T_{\rm eq} = \sqrt{r_{\rm A}} T_{\rm eff}$.

%%%%%%%%%%%%%%%%%%%%%%%%%%%%%%%%%%%%%%%%%%%%%%%%%%%%%%%%%%%%%%%%%%%%%%%%%%%%%%%%%%%%%%%%%%%%%%%%%%%%%%%%%%%%%%%%%%%%%%%%%%%%%%%%%%%%%%%%%%%%%%%%%%%%%

\section{Homogeneous Studies of Transiting Extrasolar Planets (HSTEP)} \label{sec:hstep}

Back in 2006-7 it became clear that the number of known TEPs was increasing quickly, and that studies of these objects were done in a variety of different ways, especially concerning the {\it additional constraint}. This variety of approaches led inexorably to inhomogeneous results, so the properties of different TEPs were not directly comparable. The obvious solution was an homogeneous analysis. For this, I select good published light curves and model them using the {\sc jktebop} code. Careful attention is paid to the inclusion of limb darkening, numerical integration to account for long exposure times, correction for contaminating `third' light, and in accounting for eccentric orbits. Four error analysis methods are implemented: Monte Carlo simulations, residual-permutation, multiple analyses of the same data using different choices of limb darkening, and separate analyses of different datasets for the same TEP \citep{Me08mn}.

Once the photometric parameters have been obtained, I add published spectroscopic results ($K_{\rm A}$, $T_{\rm eff}$, [Fe/H]) and calculate the physical properties of the systems. Statistical errors are prepagated from all input values by a perturbation analysis which yields a full error budget for each output value \citep{Me++05aa,Me09mn}. This process is done using each of five sets of theoretical stellar models, allowing a systematic error to be assigned to each output parameter based on the variation between the five results. Further details can be found in the original papers, and a summary has been given by \citet{Me14iaus}.

At this point, a total of 89 planetary systems have been studied in the course of HSTEP \citep{Me08mn,Me09mn,Me10mn,Me11mn,Me12mn}, most based on published data but some on new light curves obtained for the project \citep[see][and references therein]{Me+15mn}. A paper in preparation will push this number up to 120 systems.

One feature of the HSTEP results is that the error estimates for the calculated parameters often are much larger than those for published works; in many cases the results agree according to the HSTEP errorbars but not according to the published errorbars. This implies that published errorbars can be rather too small: particular offenders are CoRoT-5, CoRoT-8, CoRoT-13, Kepler-5 and Kepler-7 from Paper\,IV \citep{Me11mn}, and CoRoT-19, CoRoT-20, Kepler-15, Kepler-40 (KOI-428) and OGLE-TR-56 from Paper\,\citep{Me12mn}. Three of these systems deserve special mention.

{\bf CoRoT-8.} I found a planet radius of $1.25 \pm 0.08$\,R$_{\rm Jup}$ \citep[][sect.\,6.8]{Me11mn} versus $0.57 \pm 0.02$\,R$_{\rm Jup}$ from \citet{Borde+10aa}. The orbital ephemeris in the discovery paper is incorrect, predicting the transits in the CoRoT data to occur 0.06\,d too early.

{\bf CoRoT-13.} The CoRoT satellite obtained two light curves for this object, which strongly disagree on the transit shape. I adopted the results from the better of the two, finding a planet radius of $1.252 \pm 0.076$\,R$_{\rm Jup}$ \citep[][sect.\,6.13]{Me11mn} versus $0.885 \pm 0.014$\,R$_{\rm Jup}$ in the discovery paper \citep{Cabrera+10aa}. Whilst CoRoT-13 was thought to be an extremely dense planet with a massive core of heavy elements, my results are consistent with a typical gas giant slightly less dense than Jupiter.

{\bf OGLE-TR-56.} This was the second known TEP \citep{Konacki+03nat} and its faintness means large telescopes are required to obtain good transit light curves. \citet{Adams+11apj2} obtained many excellent light curves, and determined the physical properties of the system based on these and on an assumed $M_{\rm A}$ and $R_{\rm A}$. The problem was that the chosen $M_{\rm A}$ and $R_{\rm A}$ \citep[from][]{Torres++08apj} were inconsistent with the $\rho_{\rm A}$ from the light curve. The HSTEP analysis changed the measured planetary radius from $1.378 \pm 0.090$\,R$_{\rm Jup}$ to $1.734 \pm 0.061$\,R$_{\rm Jup}$ \citep[][sect.\,5.21]{Me12mn}.

By far the most common method of obtaining errorbars on measured parameters of TEPs and their host stars is that of MCMC (Markov chain Monte Carlo), a very powerful technique for both model optimisation and calculation of the posterior probability density for parameter values. A common feature of the HSTEP reanalysis of published data is agreement with published results within the HSTEP errorbars but not with the often very small errorbars calculated using MCMC in these publications. This suggests that the error analysis methods using the HSTEP project are robust, but that those arising from MCMC analysis sometimes are not. Like any other statistical tool, MCMC has to be used carefully to ensure good results.

\subsection{TEPCat: the catalogue of physical properties of transiting extrasolar planets}

\begin{figure}[t]
\includegraphics[width=\textwidth,angle=0]{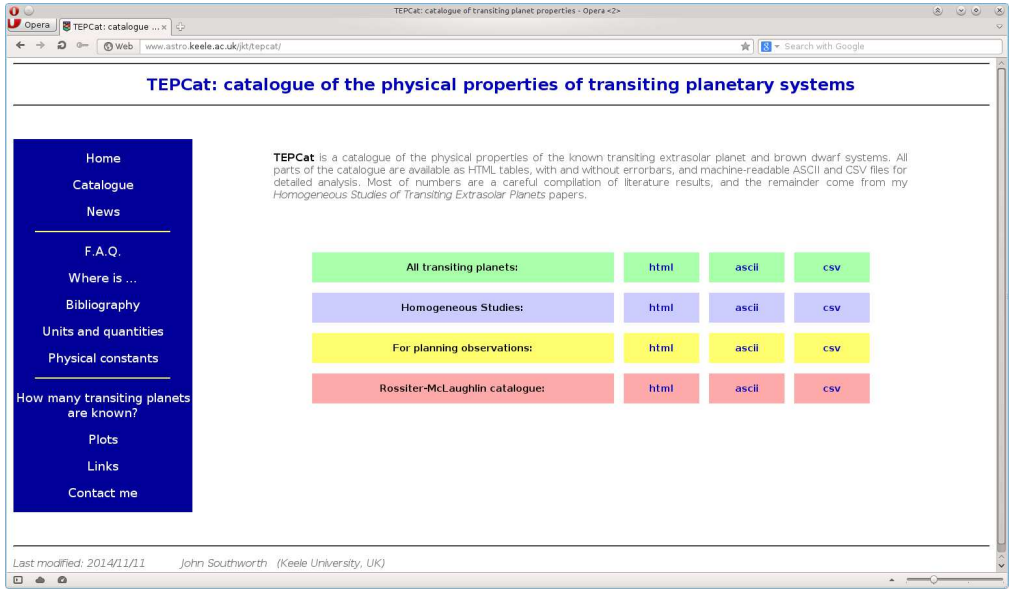} \\[-15pt]
\caption{\label{fig:tepcat} Homepage of the TEPCat catalogue
of transiting extrasolar planets.} \end{figure}

By Paper\,IV \citep{Me11mn} it was obvious that readers could not reasonably be expected to trawl through all four papers to compile the full results from the HSTEP project. I therefore created the TEPCat catalogue\footnote{TEPCat \citep{Me11mn} can be found at: {\tt http://www.astro.keele.ac.uk/jkt/tepcat/}} to make these results available in convenient formats (Fig.\,\ref{fig:tepcat}). It was also a good site for placing a compilation of the physical properties of {\em all} known TEPs and their host stars, a database which I was already keeping for my own use. At this point TEPCat contains tables in {\tt html}, {\tt ascii} and {\tt csv} formats with the best available values for the stellar properties ($T_{\rm eff}$, [Fe/H], $M_{\rm A}$, $R_{\rm A}$, $\log g$, $\rho_{\rm A}$), planet characteristics ($T_{\rm eq}$, $M_{\rm b}$, $R_{\rm b}$, $g_{\rm b}$, $\rho_{\rm b}$), orbital parameters ($P$, $T_0$, $e$, $a$) and references for all confirmed TEPs. A catalogue of orbital obliquities from the Rossiter-McLaughlin effect is also maintained, along with various goodies such as plots, links, explanation, and the set of physical constants used in the HSTEP project.

%%%%%%%%%%%%%%%%%%%%%%%%%%%%%%%%%%%%%%%%%%%%%%%%%%%%%%%%%%%%%%%%%%%%%%%%%%%%%%%%%%%%%%%%%%%%%%%%%%%%%%%%%%%%%%%%%%%%%%%%%%%%%%%%%%%%%%%%%%%%%%%%%%%%%

\section{Rossiter-McLaughlin effect}

\begin{figure}[t]
\includegraphics[width=\textwidth,angle=0]{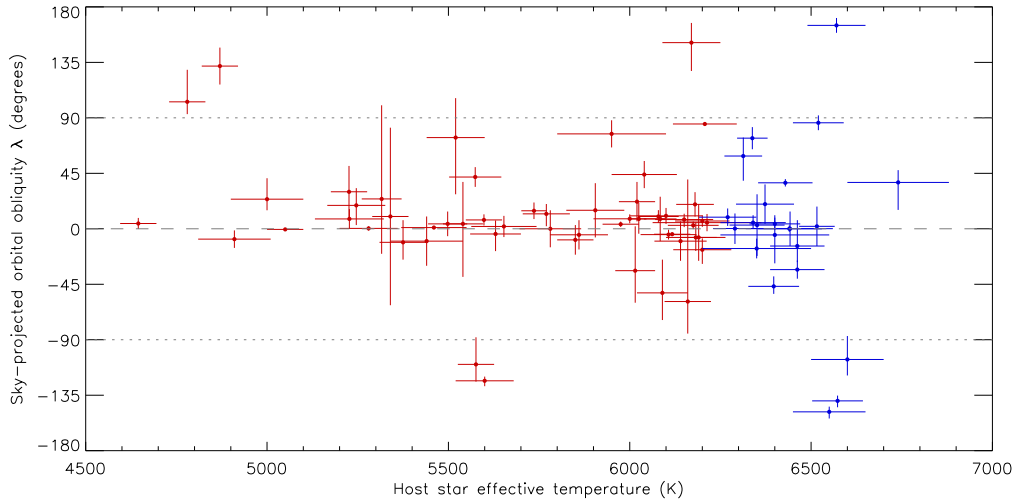} \\[-15pt]
\caption{\label{fig:rm} Sky-projected orbital obliquity measurements $\lambda$. Red and
blue are used for systems where the host star is cooler or hotter than 6250\,K. The grey
dashed line indicates perfect alignment and the grey dotted lines show the $\pm$90$^\circ$
boundaries outside which orbits are retrograde. Data were taken from TEPCat.} \end{figure}

\citet{Rossiter24apj} and \citet{Mclaughlin24apj} contemporaneously discovered an RV anomaly during primary eclipse, in the eclipsing binaries $\beta$\,Lyrae and $\beta$\,Persei. This is caused by the eclipsing object blocking out part of the rotating surface of its companion, removing flux from part of its spectral line profiles and thus biasing measured RVs away from the Keplerian value. The effect is much smaller in TEPs (typically less than 50\,m\,s$^{-1}$ versus 13\,km\,s$^{-1}$ for $\beta$\,Lyrae) but easier to study because the spectral line profiles come from only one object (the host star). The Rossiter-McLaughlin (RM) effect has now been observed in a total of 91 TEPs \citep[e.g.][]{Triaud+10aa,Albrecht+12apj2}, mostly by RV measurements. This approach can only give the sky-projected value ($\lambda$) of the true orbital obliquity ($\psi$).

Whilst early RM measurements \citep[the first being][]{Queloz+00aa} indicated aligned orbits, a significant number of misaligned and even retrograde planets are now known \citep[the first being WASP-17;][]{Anderson+10apj}. \citet{Winn+10apj3} found that misaligned orbits occur mostly for hotter host stars ($T_{\rm eff} > 6250$\,K), although \citet{Triaud11aa} asserted that this was caused by the younger age of such systems. Tidal dissipation is a critical part of interpreting $\lambda$ measurements \citep[see][]{Albrecht+12apj2}.

%%%%%%%%%%%%%%%%%%%%%%%%%%%%%%%%%%%%%%%%%%%%%%%%%%%%%%%%%%%%%%%%%%%%%%%%%%%%%%%%%%%%%%%%%%%%%%%%%%%%%%%%%%%%%%%%%%%%%%%%%%%%%%%%%%%%%%%%%%%%%%%%%%%%%

\section{Starspots}

\begin{figure}[t]
\includegraphics[width=\textwidth,angle=0]{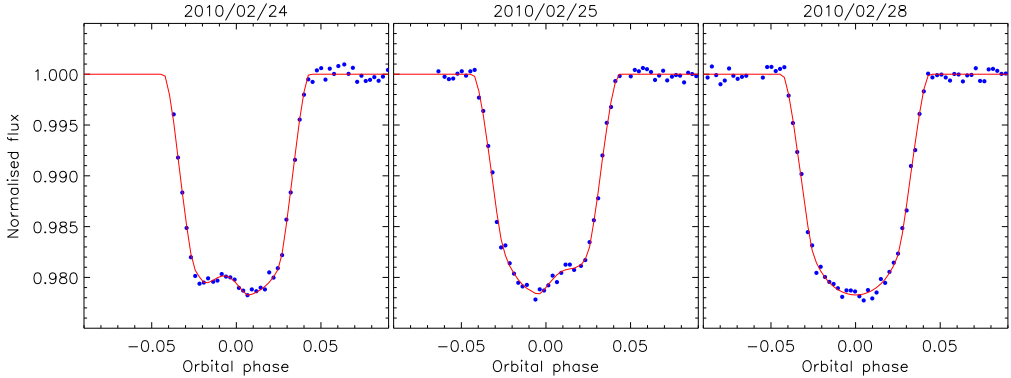} \\[-15pt]
\caption{\label{fig:wasp19} Three transits of WASP-19 taken over five
nights by \citet{Tregloan++13mn}. Starspot anomalies are visible in
the first two transits.} \end{figure}

An alternative way to measure the RM effect is via transits with starspot anomalies. If a planet transits a dark spot on the stellar surface, it temporarily blocks slightly less of the overall starlight. The overall brightness of the system blips upwards, by an amount which depends on the size of the spot and its brightness relative to the rest of the stellar surface. Multi-band photometry of this effect allows the spot temperature to be obtained \citep[e.g.][]{Mancini+14mn} and the spot position to be measured precisely.

If several transits are observed over a short period of time, the change in position of a starspot could be tracked. This directly yields the motion of the spot as the star rotates, relative to the planet's orbit, allowing $\lambda$ to be measured \citep{Nutzman+11apj2,Sanchis+11apj} as well as the rotation period of the star \citep{Silva08apj}. \citet{Tregloan++13mn} constructed a physically realistic model of this situation ({\sc prism}) and used it to measure $\lambda = 1.0^\circ \pm 1.2^\circ$ from two transits of WASP-19 (Fig.\,\ref{fig:wasp19}), a much more precise value than the RM alternative of $\lambda = 4.6^\circ \pm 5.2^\circ$ \citep{Hellier+11apj}. Having three or more observations of the same starspot at different positions would allow $\psi$ as well as $\lambda$ to be obtained.

%%%%%%%%%%%%%%%%%%%%%%%%%%%%%%%%%%%%%%%%%%%%%%%%%%%%%%%%%%%%%%%%%%%%%%%%%%%%%%%%%%%%%%%%%%%%%%%%%%%%%%%%%%%%%%%%%%%%%%%%%%%%%%%%%%%%%%%%%%%%%%%%%%%%%

\section{Occultations}

Although planets are much fainter than their host stars, it is possible to detect the dips in brightness as they are eclipsed by their star. These miniscule occultations can only be measured using very high-precision photometry, but are nevertheless a valuable source of two types of information.

Firstly, the time of mid-occultation constrains $e$ and $\omega$ for eccentric systems. Specifically, the difference in orbital phase between the occultation and the halfway point between the two adjacent transits gives the combination term $e\cos\omega$ independently of RV measurements:
$$ \Delta\phi = \left(\frac{t_{\rm occult}-t_{\rm transit}}{P}\right) - 0.5 = \left(\frac{1+\csc^2i}{\pi}\right) e\cos\omega $$
Here, $\phi$ means orbital phase, and $t_{\rm transit}$ and $t_{\rm occult}$ indicate the eclipse midpoints

Secondly, the depth of the occultation gives the brightness of the planet (at a given wavelength or passband) relative to that of the star. This means that the spectrum of the planet can be constructed from occultation observations at a range of wavelengths. The spectrum is that of the irradiated `dayside' of the planet. As a rough approximation, reflected light from the star dominates at optical wavelengths \citep[e.g.][]{Angerhausen++14xxx} and thermal emission dominates in the infrared \citep[e.g.][]{Charbonneau+08apj}. Planetary spectra can be used to investigate the chemical composition and structure of their atmospheres \citep[e.g.][]{Ranjan+14apj,MadhusudhanSeager10apj}.

%%%%%%%%%%%%%%%%%%%%%%%%%%%%%%%%%%%%%%%%%%%%%%%%%%%%%%%%%%%%%%%%%%%%%%%%%%%%%%%%%%%%%%%%%%%%%%%%%%%%%%%%%%%%%%%%%%%%%%%%%%%%%%%%%%%%%%%%%%%%%%%%%%%%%

\section{Transmission spectroscopy}

Whilst occultations can be used to measure the flux emitted by the planet, observations of the depth of the transit as a function of wavelength allow the opacity spectrum of a planetary atmosphere to be measured. Occultations probe the dayside of the planet whereas transmission spectroscopy is sensitive to the properties of the atmosphere at the terminator (the limb of the planet). This effect is difficult to observe, but is helped by the very extended atmospheres of some planets. The best example is WASP-17 \citep{Anderson+10apj}, which is the largest known planet at $R_{\rm b} = 1.932 \pm 0.053$\,R$_{\rm Jup}$ \citep{Me+12mn3}. Its low surface gravity of $g_{\rm b} = 3.16 \pm 0.20$\,m\,s$^{-1}$ yields a huge atmospheric scale height of 2000\,km (0.028\,R$_{\rm Jup}$). The largest features in the optical and near-infrared spectrum of a hot Jupiter can be 5--10 atmospheric scale heights \citep{Sing+11mn,Sing+14xxx}, which are detectable using ground-based large telescopes.

Theoretical spectra of irradiated giant planets show characteristic features at optical wavelengths due to sodium and potassium \citep{Fortney+08apj}, possibly sulphur compounds \citep{Zahnle+09apj}, and Rayleigh and Mie scattering in the blue. Infrared wavelengths are predicted to show features due to molecules such as H$_2$O, CO, CO$_2$ and CH$_4$, depending on the atmospheric temperature. This is an active area of research which has generated a wide variety of results: some planets show flat transmission spectra indicative of high-altitude clouds \citep{Kreidberg+14nat,Knutson+14nat}, some show signatures of molecules \citep{Tinetti+10apj,Wakeford+13mn}, some show Rayleigh or Mie scattering \citep{Pont+13mn,Sing+13mn}, and at least one planet shows all of these features \citep{Sing+14xxx}.

%%%%%%%%%%%%%%%%%%%%%%%%%%%%%%%%%%%%%%%%%%%%%%%%%%%%%%%%%%%%%%%%%%%%%%%%%%%%%%%%%%%%%%%%%%%%%%%%%%%%%%%%%%%%%%%%%%%%%%%%%%%%%%%%%%%%%%%%%%%%%%%%%%%%%

\section{Future}

We have passed through the initial stages of development of the study of transiting planets and are now in the early characterisation phase. The easy-to-find TEPs are being identified in bulk by ground-based surveys \citep[e.g.][]{Bakos+12aj,Hellier+12mn} and our boundaries of ignorance are being gradually pushed back by discoveries in new areas of parameter space \citep[e.g.][]{Doyle+11sci,Sanchis+13apj,Ciceri+14xxx}. Exhaustive examinations of a small subset of TEPs have established them as tracers of the formation, structure and evolution of giant planets. The best-studied TEPs have mass and radius measurements to a few percent precision, projected or true orbital obliquity measurements, and atmospheric abundances of some atoms and molecules through transit and occultation spectroscopy. Whilst {\it Kepler} has truly revolutionised the study of TEPs, ground-based surveys remain relevant as they observe many more targets so can find rarer types of planet.

In the near future, the {\it Gaia} satellite will fill an important hole in our understanding of TEPs and their host stars. The {\it Gaia} parallax measurements will give direct distance and thus luminosity estimates. As $L = 4 \pi R^2 T_{\rm eff}^{\,4}$, these parallaxes can replace the {\it additional constraint} which troubles existing mass and radius measurements of TEPs (see Sect.\,\ref{sec:hstep}). The high photometric precision of {\it Gaia} will also enable it to be used to discover TEPs \citep[see][]{DziganZucker12apj}, although it is likely that huge observational resources would be needed to follow up the identified planet candidates.

Although the main missions of {\it Kepler} and CoRoT have been terminated by technical problems, their archives remain rich in untapped results. {\it Kepler} has been reincarnated as the K2 mission, with lower photometric precision but still much better than achievable from the ground. CHEOPS \citep{Broeg+13conf} is slated for launch in 2017, for a 3.5\,yr mission to detect transits in low-mass planets discovered by the RV method.

The next landmark mission is TESS \citep{Ricker+14spie}, also due for launch in 2017. TESS will photometrically observe 26 fields covering most of the sky, concentrating on bright stars but for much shorter time intervals (27 days near the ecliptic ranging to one year around the celestial poles). Further ahead, the PLATO mission \citep{Rauer+14xxx} is planned for launch in 2024 as a precision photometry survey instrument. PLATO will have a much larger field of view than {\it Kepler}: it will observe brighter stars which makes follow-up observations much easier. It will also observe patches of sky for several years, thereby avoiding the low sensitivity to long-period planets suffered by TESS. Our knowledge of transiting planets is set to improve immensely.

%%%%%%%%%%%%%%%%%%%%%%%%%%%%%%%%%%%%%%%%%%%%%%%%%%%%%%%%%%%%%%%%%%%%%%%%%%%%%%%%%%%%%%%%%%%%%%%%%%%%%%%%%%%%%%%%%%%%%%%%%%%%%%%%%%%%%%%%%%%%%%%%%%%%%

% \bibliography{jkt}

\begin{thebibliography}{}
\expandafter\ifx\csname natexlab\endcsname\relax\def\natexlab#1{#1}\fi
\expandafter\ifx\csname url\endcsname\relax
  \def\url#1{\texttt{#1}}\fi
\expandafter\ifx\csname urlprefix\endcsname\relax\def\urlprefix{URL }\fi
\providecommand{\eprint}[2][]{\url{#2}}

\bibitem[{{Adams} et~al.(2011){Adams}, {L{\'o}pez-Morales}, {Elliot}, {Seager},
  {Osip}, {Holman}, {Winn}, {Hoyer}, \& {Rojo}}]{Adams+11apj2}
{Adams}, E.~R., et al. 2011, ApJ, 741, 102

\bibitem[{{Albrecht} et~al.(2012){Albrecht}, {Winn}, {Johnson}, {Howard},
  {Marcy}, {Butler}, {Arriagada}, {Crane}, {Shectman}, {Thompson}, {Hirano},
  {Bakos}, \& {Hartman}}]{Albrecht+12apj2}
{Albrecht}, S., et al. 2012, ApJ, 757, 18

\bibitem[{{Alonso} et~al.(2008){Alonso}, {Barbieri}, {Rabus}, {Deeg},
  {Belmonte}, \& {Almenara}}]{Alonso+08aa}
{Alonso}, R., et al. 2008, A\&A, 487, L5

\bibitem[{{Anderson} et~al.(2010){Anderson}, {Hellier}, {Gillon}, {Triaud},
  {Smalley}, {Hebb}, {Collier Cameron}, {Maxted}, {Queloz}, {West}, \&
  {Others}}]{Anderson+10apj}
{Anderson}, D.~R., et al. 2010, ApJ, 709, 159

\bibitem[{{Angerhausen} et~al.(2014){Angerhausen}, {DeLarme}, \&
  {Morse}}]{Angerhausen++14xxx}
{Angerhausen}, D., {DeLarme}, E., \& {Morse}, J.~A. 2014, {\tt arXiv:1404.4348}

\bibitem[{{Bakos} et~al.(2012){Bakos}, {Hartman}, {Torres}, {B{\'e}ky},
  {Latham}, {Buchhave}, {Csubry}, {Kov{\'a}cs}, {Bieryla}, {Quinn}, \&
  {Others}}]{Bakos+12aj}
{Bakos}, G.~{\'A}., et al. 2012, AJ, 144, 19

\bibitem[{{Bakos} et~al.(2002){Bakos}, {L{\'a}z{\'a}r}, {Papp}, {S{\'a}ri}, \&
  {Green}}]{Bakos+02pasp}
{Bakos}, G.~{\'A}., {L{\'a}z{\'a}r}, J., {Papp}, I., {S{\'a}ri}, P., \&
  {Green}, E.~M. 2002, PASP, 114, 974

\bibitem[{{Bord{\'e}} et~al.(2010){Bord{\'e}}, {Bouchy}, {Deleuil}, {Cabrera},
  {Jorda}, {Lovis}, {Csizmadia}, {Aigrain}, {Almenara}, {Alonso}, \&
  {Others}}]{Borde+10aa}
{Bord{\'e}}, P., et al. 2010, A\&A, 520, A66

\bibitem[{{Borucki} et~al.(2010){Borucki}, {Koch}, {Basri}, {Batalha}, {Brown},
  {Caldwell}, {Caldwell}, {Christensen-Dalsgaard}, {Cochran}, \&
  {Others}}]{Borucki+10sci}
{Borucki}, W.~J., et al. 2010, Science, 327, 977

\bibitem[{{Bouchy} et~al.(2005){Bouchy}, {Udry}, {Mayor}, {Moutou}, {Pont},
  {Iribarne}, {da Silva}, {Ilovaisky}, {Queloz}, {Santos}, {S{\'e}gransan}, \&
  {Zucker}}]{Bouchy+05aa}
{Bouchy}, F., et al. 2005, A\&A, 444, L15

\bibitem[{{Broeg} et~al.(2013){Broeg}, {Fortier}, {Ehrenreich}, {Alibert},
  {Baumjohann}, {Benz}, {Deleuil}, {Gillon}, {Ivanov}, {Liseau}, \&
  {Others}}]{Broeg+13conf}
{Broeg}, C., et al. 2013, in European Physical Journal Web of Conferences, vol.~47, 3005

\bibitem[{{Cabrera} et~al.(2010){Cabrera}, {Bruntt}, {Ollivier}, {Diaz},
  {Csizmadia}, {Aigrain}, {Alonso}, {Almenara}, {Auvergne}, {Baglin}, \&
  {Others}}]{Cabrera+10aa}
{Cabrera}, J., et al. 2010, A\&A, 522, A110

\bibitem[{{Charbonneau} et~al.(2000){Charbonneau}, {Brown}, {Latham}, \&
  {Mayor}}]{Charbonneau+00apj}
{Charbonneau}, D., {Brown}, T.~M., {Latham}, D.~W., \& {Mayor}, M. 2000, ApJ, 529, L45

\bibitem[{{Charbonneau} et~al.(2008){Charbonneau}, {Knutson}, {Barman},
  {Allen}, {Mayor}, {Megeath}, {Queloz}, \& {Udry}}]{Charbonneau+08apj}
{Charbonneau}, D., et al. 2008, ApJ, 686, 1341

\bibitem[{{Charbonneau} et~al.(2007){Charbonneau}, {Winn}, {Everett}, {Latham},
  {Holman}, {Esquerdo}, \& {O'Donovan}}]{Charbonneau+07apj}
{Charbonneau}, D., et al. 2007, ApJ, 658, 1322

\bibitem[{{Ciceri} et~al.(2014){Ciceri}, {Lillo-Box}, {Southworth}, {Mancini},
  {Henning}, \& {Barrado}}]{Ciceri+14xxx}
{Ciceri}, S., et al. 2014, A\&A, submitted, {\tt arXiv1410.2999}

\bibitem[{{Claret} \& {Bloemen}(2011)}]{ClaretBloemen11aa}
{Claret}, A., \& {Bloemen}, S. 2011, A\&A, 529, A75

\bibitem[{{Collier Cameron} et~al.(2007){Collier Cameron}, {Bouchy},
  {H{\'e}brard}, {Maxted}, {Pollacco}, {Pont}, {Skillen}, {Smalley}, {Street},
  {West}, {Wilson}, \& {Others}}]{Cameron+07mn}
{Collier Cameron}, A., et al. 2007, MNRAS, 375, 951

\bibitem[{{Coughlin} et~al.(2014){Coughlin}, {Thompson}, {Bryson}, {Burke},
  {Caldwell}, {Christiansen}, {Haas}, {Howell}, {Jenkins}, {Kolodziejczak},
  {Mullally}, \& {Rowe}}]{Coughlin+14aj}
{Coughlin}, J.~L., et al. 2014, AJ, 147, 119

\bibitem[{{Daemgen} et~al.(2009){Daemgen}, {Hormuth}, {Brandner}, {Bergfors},
  {Janson}, {Hippler}, \& {Henning}}]{Daemgen+09aa}
{Daemgen}, S., et al. 2009, A\&A, 498, 567

\bibitem[{{Doyle} et~al.(2011){Doyle}, {Carter}, {Fabrycky}, {Slawson},
  {Howell}, {Winn}, {Orosz}, {Prsa}, {Welsh}, {Quinn}, \&
  {Others}}]{Doyle+11sci}
{Doyle}, L.~R., et al. 2011, Science, 333, 1602

\bibitem[{{Dzigan} \& {Zucker}(2012)}]{DziganZucker12apj}
{Dzigan}, Y., \& {Zucker}, S. 2012, ApJ, 753, L1

\bibitem[{{Enoch} et~al.(2010){Enoch}, {Collier Cameron}, {Parley}, \&
  {Hebb}}]{Enoch+10aa}
{Enoch}, B., {Collier Cameron}, A., {Parley}, N.~R., \& {Hebb}, L. 2010, A\&A, 516, A33

\bibitem[{{Fortney} et~al.(2008){Fortney}, {Lodders}, {Marley}, \&
  {Freedman}}]{Fortney+08apj}
{Fortney}, J.~J., {Lodders}, K., {Marley}, M.~S., \& {Freedman}, R.~S. 2008, ApJ, 678, 1419

\bibitem[{{Hellier} et~al.(2012){Hellier}, {Anderson}, {Collier Cameron},
  {Doyle}, {Fumel}, {Gillon}, {Jehin}, {Lendl}, {Maxted}, {Pepe}, \&
  {Others}}]{Hellier+12mn}
{Hellier}, C., et al. 2012, MNRAS, 426, 739

\bibitem[{{Hellier} et~al.(2011{\natexlab{a}}){Hellier}, {Anderson}, {Collier
  Cameron}, {Gillon}, {Lendl}, {Lister}, {Maxted}, {Pollacco}, {Queloz},
  {Smalley}, {Triaud}, \& {West}}]{Hellier+11conf}
{Hellier}, C., et al. 2011{\natexlab{a}}, in
  European Physical Journal Web of Conferences, vol.~11, 1004

\bibitem[{{Hellier} et~al.(2011{\natexlab{b}}){Hellier}, {Anderson},
  {Collier-Cameron}, {Miller}, {Queloz}, {Smalley}, {Southworth}, \&
  {Triaud}}]{Hellier+11apj}
{Hellier}, C., et al. 2011{\natexlab{b}}, ApJ, 730, L31

\bibitem[{{Henry} et~al.(2000){Henry}, {Marcy}, {Butler}, \&
  {Vogt}}]{Henry+00apj}
{Henry}, G.~W., {Marcy}, G.~W., {Butler}, R.~P., \& {Vogt}, S.~S. 2000, ApJ, 529, L41

\bibitem[{{Hilditch}(2001)}]{Hilditch01book}
{Hilditch}, R.~W. 2001, {An Introduction to Close Binary Stars} (Cambridge University Press, Cambridge, UK)

\bibitem[{{Jenkins} et~al.(2010){Jenkins}, {Caldwell}, {Chandrasekaran},
  {Twicken}, {Bryson}, {Quintana}, {Clarke}, {Li}, {Allen}, {Tenenbaum}, \&
  {Others}}]{Jenkins+10apj}
{Jenkins}, J.~M., et al. 2010, ApJ, 713, L120

\bibitem[{{Kipping}(2008)}]{Kipping08mn}
{Kipping}, D.~M. 2008, MNRAS, 389, 1383

\bibitem[{{Knutson} et~al.(2014){Knutson}, {Benneke}, {Deming}, \&
  {Homeier}}]{Knutson+14nat}
{Knutson}, H.~A., {Benneke}, B., {Deming}, D., \& {Homeier}, D. 2014, Nature, 505, 66

\bibitem[{{Konacki} et~al.(2003){Konacki}, {Torres}, {Jha}, \&
  {Sasselov}}]{Konacki+03nat}
{Konacki}, M., {Torres}, G., {Jha}, S., \& {Sasselov}, D.~D. 2003, Nature, 421, 507

\bibitem[{{Kreidberg} et~al.(2014){Kreidberg}, {Bean}, {D{\'e}sert}, {Benneke},
  {Deming}, {Stevenson}, {Seager}, {Berta-Thompson}, {Seifahrt}, \&
  {Homeier}}]{Kreidberg+14nat}
{Kreidberg}, L., et al. 2014, Nature, 505, 69

\bibitem[{{Lillo-Box} et~al.(2014){Lillo-Box}, {Barrado}, \&
  {Bouy}}]{Lillobox++14aa}
{Lillo-Box}, J., {Barrado}, D., \& {Bouy}, H. 2014, A\&A, in press, {\tt arXiv:1405.3120}

\bibitem[{{Madhusudhan} \& {Seager}(2010)}]{MadhusudhanSeager10apj}
{Madhusudhan}, N., \& {Seager}, S. 2010, ApJ, 725, 261

\bibitem[{{Mancini} et~al.(2014){Mancini}, {Southworth}, {Ciceri},
  {Tregloan-Reed}, {Crossfield}, {Nikolov}, {Bruni}, {Zambelli}, \&
  {Henning}}]{Mancini+14mn}
{Mancini}, L., et al. 2014, MNRAS, 443, 2391

\bibitem[{{Mandel} \& {Agol}(2002)}]{MandelAgol02apj}
{Mandel}, K., \& {Agol}, E. 2002, ApJ, 580, L171

\bibitem[{{Marcy} \& {Butler}(1996)}]{MarcyButler96apj}
{Marcy}, G.~W., \& {Butler}, R.~P. 1996, ApJ, 464, L147

\bibitem[{{Masuda}(2014)}]{Masuda14apj}
{Masuda}, K. 2014, ApJ, 783, 53

\bibitem[{{Mayor} \& {Queloz}(1995)}]{MayorQueloz95nat}
{Mayor}, M., \& {Queloz}, D. 1995, Nature, 378, 355

\bibitem[{{McLaughlin}(1924)}]{Mclaughlin24apj}
{McLaughlin}, D.~B. 1924, ApJ, 60, 22

\bibitem[{{Morton} \& {Johnson}(2011)}]{MortonJohnson11apj}
{Morton}, T.~D., \& {Johnson}, J.~A. 2011, ApJ, 738, 170

\bibitem[{{Moutou} et~al.(2013){Moutou}, {Deleuil}, {Guillot}, {Baglin},
  {Bord{\'e}}, {Bouchy}, {Cabrera}, {Csizmadia}, \& {Deeg}}]{Moutou+13icar}
{Moutou}, C., et al. 2013, Icarus, 226, 1625

\bibitem[{{Nutzman} et~al.(2011){Nutzman}, {Fabrycky}, \&
  {Fortney}}]{Nutzman+11apj2}
{Nutzman}, P.~A., {Fabrycky}, D.~C., \& {Fortney}, J.~J. 2011, ApJ, 740, L10

\bibitem[{{Pepe} et~al.(2014){Pepe}, {Ehrenreich}, \& {Meyer}}]{Pepe+14nat}
{Pepe}, F., {Ehrenreich}, D., \& {Meyer}, M.~R. 2014, Nature, 513, 358

\bibitem[{{Pollacco} et~al.(2006){Pollacco}, {Skillen}, {Cameron}, {Christian},
  {Hellier}, {Irwin}, {Lister}, {Street}, {West}, \&
  {Others}}]{Pollacco+06pasp}
{Pollacco}, D.~L., et al. 2006, PASP, 118, 1407

\bibitem[{{Pont} et~al.(2013){Pont}, {Sing}, {Gibson}, {Aigrain}, {Henry}, \&
  {Husnoo}}]{Pont+13mn}
{Pont}, F., et al. 2013, MNRAS, 432, 2917

\bibitem[{{Queloz} et~al.(2000){Queloz}, {Eggenberger}, {Mayor}, {Perrier},
  {Beuzit}, {Naef}, {Sivan}, \& {Udry}}]{Queloz+00aa}
{Queloz}, et al. 2000, A\&A, 359, L13

\bibitem[{{Ranjan} et~al.(2014){Ranjan}, {Charbonneau}, {D{\'e}sert},
  {Madhusudhan}, {Deming}, {Wilkins}, \& {Mandell}}]{Ranjan+14apj}
{Ranjan}, S., et al. 2014, ApJ, 785, 148

\bibitem[{{Rauer} et~al.(2014){Rauer}, {Catala}, {Aerts}, {Appourchaux},
  {Benz}, {Brandeker}, {Christensen-Dalsgaard}, {Deleuil}, {Gizon}, {Goupil},
  \& {Others}}]{Rauer+14xxx}
{Rauer}, H., et al. 2014, Experimental Astronomy, in press, {\tt arXiv:1310.0696}

\bibitem[{{Ricker} et~al.(2014){Ricker}, {Winn}, {Vanderspek}, {Latham},
  {Bakos}, {Bean}, {Berta-Thompson}, {Brown}, {Buchhave}, {Butler}, \&
  {Others}}]{Ricker+14spie}
{Ricker}, G.~R., et al. 2014, in SPIE Conference Series, vol.~9143, 20

\bibitem[{{Rossiter}(1924)}]{Rossiter24apj}
{Rossiter}, R.~A. 1924, ApJ, 60, 15

\bibitem[{{Rowe} et~al.(2014){Rowe}, {Bryson}, {Marcy}, {Lissauer},
  {Jontof-Hutter}, {Mullally}, {Gilliland}, {Issacson}, {Ford}, {Howell}, \&
  {Others}}]{Rowe+14apj}
{Rowe}, J.~F., et al. 2014, ApJ, 784, 45

\bibitem[{{Sanchis-Ojeda} et~al.(2011){Sanchis-Ojeda}, {Winn}, {Holman},
  {Carter}, {Osip}, \& {Fuentes}}]{Sanchis+11apj}
{Sanchis-Ojeda}, R., et al. 2011, ApJ, 733, 127

\bibitem[{{Sanchis-Ojeda} et~al.(2013){Sanchis-Ojeda}, {Rappaport}, {Winn},
  {Levine}, {Kotson}, {Latham}, \& {Buchhave}}]{Sanchis+13apj}
{Sanchis-Ojeda}, R., et al. 2013, ApJ, 774, 54

\bibitem[{{Santerne} et~al.(2012){Santerne}, {D{\'{\i}}az}, {Moutou}, {Bouchy},
  {H{\'e}brard}, {Almenara}, {Bonomo}, {Deleuil}, \& {Santos}}]{Santerne+12aa}
{Santerne}, A., et al. 2012, A\&A, 545, A76

\bibitem[{{Seager} \& {Mall{\'e}n-Ornelas}(2003)}]{SeagerMallen03apj}
{Seager}, S., \& {Mall{\'e}n-Ornelas}, G. 2003, ApJ, 585, 1038

\bibitem[{{Sheets} \& {Deming}(2014)}]{SheetsDeming14apj}
{Sheets}, H.~A., \& {Deming}, D. 2014, ApJ, 794, 133

\bibitem[{{Silva-Valio}(2008)}]{Silva08apj}
{Silva-Valio}, A. 2008, ApJ, 683, L179

\bibitem[{{Sing} et~al.(2011){Sing}, {Pont}, {Aigrain}, {Charbonneau},
  {D{\'e}sert}, {Gibson}, {Gilliland}, {Hayek}, {Henry}, {Knutson}, {Lecavelier
  Des Etangs}, {Mazeh}, \& {Shporer}}]{Sing+11mn}
{Sing}, D.~K., et al. 2011, MNRAS, 416, 1443

\bibitem[{{Sing} et~al.(2013){Sing}, {Lecavelier des Etangs}, {Fortney},
  {Burrows}, {Pont}, {Wakeford}, {Ballester}, {Nikolov}, {Henry}, {Aigrain}, \&
  {Others}}]{Sing+13mn}
{Sing}, D.~K., et al. 2013, MNRAS, 436, 2956

\bibitem[{{Sing} et~al.(2014){Sing}, {Wakeford}, {Showman}, {Nikolov},
  {Fortney}, {Burrows}, {Ballester}, {Deming}, {Aigrain}, {D{\'e}sert},
  {Gibson}, {Henry}, {Knutson}, {Lecavelier des Etangs}, {Pont},
  {Vidal-Madjar}, {Williamson}, \& {Wilson}}]{Sing+14xxx}
{Sing}, D.~K., et al. 2014, MNRAS, in press, {\tt arXiv:1410.7611}

\bibitem[{{Southworth}(2008)}]{Me08mn}
{Southworth}, J. 2008, MNRAS, 386, 1644

\bibitem[{{Southworth}(2009)}]{Me09mn}
--- 2009, MNRAS, 394, 272

\bibitem[{{Southworth}(2010)}]{Me10mn}
--- 2010, MNRAS, 408, 1689

\bibitem[{{Southworth}(2011)}]{Me11mn}
--- 2011, MNRAS, 417, 2166

\bibitem[{{Southworth}(2012)}]{Me12mn}
--- 2012, MNRAS, 426, 1291

\bibitem[{{Southworth}(2013)}]{Me13aa}
--- 2013, A\&A, 557, A119

\bibitem[{{Southworth}(2014{\natexlab{a}})}]{Me14iaus}
--- 2014{\natexlab{a}}, in IAU Symposium 293, edited by N.~{Haghighipour}, 423

\bibitem[{{Southworth}(2014{\natexlab{b}})}]{Me14debcat}
--- 2014{\natexlab{b}}, EPJ Web of Science, in press, {\tt arXiv:1411.1219}

\bibitem[{{Southworth} et~al.(2012){Southworth}, {Hinse}, {Dominik}, {Fang},
  {Harps{\o}e}, {J{\o}rgensen}, {Kerins}, {Liebig}, {Mancini}, {Skottfelt},
  {Anderson}, {Smalley}, {Tregloan-Reed}, {Wertz}, {Alsubai}, {Bozza}, {Calchi
  Novati}, {Dreizler}, {Gu}, {Hundertmark}, {Jessen-Hansen}, {Kains},
  {Kjeldsen}, {Lund}, {Lundkvist}, {Mathiasen}, {Penny}, {Rahvar}, {Ricci},
  {Scarpetta}, {Snodgrass}, \& {Surdej}}]{Me+12mn3}
{Southworth}, J., et al. 2012, MNRAS, 426, 1338

\bibitem[{{Southworth} et~al.(2009){Southworth}, {Hinse}, {J{\o}rgensen},
  {Dominik}, {Ricci}, {Burgdorf}, {Hornstrup}, {Wheatley}, {Anguita}, {Bozza},
  {Calchi Novati}, {Harps{\o}e}, {Kj{\ae}rgaard}, {Liebig}, {Mancini}, {Masi},
  {Mathiasen}, {Rahvar}, {Scarpetta}, {Snodgrass}, {Surdej}, {Th{\"o}ne}, \&
  {Zub}}]{Me+09mn}
{Southworth}, J., et al. 2009, MNRAS, 396, 1023

\bibitem[{{Southworth} et~al.(2010){Southworth}, {Mancini}, {Calchi Novati},
  {Dominik}, {Glitrup}, {Hinse}, {J{\o}rgensen}, {Mathiasen}, {Ricci}, {Maier},
  {Zimmer}, {Bozza}, {Browne}, {Bruni}, {Burgdorf}, {Dall'Ora}, {Finet},
  {Harps{\o}e}, {Hundertmark}, {Liebig}, {Rahvar}, {Scarpetta}, {Skottfelt},
  {Smalley}, {Snodgrass}, \& {Surdej}}]{Me+10mn}
{Southworth}, J., et al. 2010, MNRAS, 408, 1680

\bibitem[{{Southworth} et~al.(2014){Southworth}, {Mancini}, {Ciceri}, {Budaj},
  {Dominik}, {Figuera Jaimes}, {Haugbolle}, {Jorgensen}, {Popovas}, {Rabus},
  {Rahvar}, {von Essen}, {Schmidt}, {Wertz}, {Alsubai}, {Bozza}, {Bramich},
  {Calchi Novati}, {D'Ago}, {Hinse}, {Henning}, {Hundertmark}, {Juncher},
  {Korhonen}, {Skottfelt}, {Snodgrass}, {Starkey}, \& {Surdej}}]{Me+15mn}
{Southworth}, J., et al. 2014, MNRAS, in press, {\tt arXiv:1411.2767}

\bibitem[{{Southworth} et~al.(2005){Southworth}, {Maxted}, \&
  {Smalley}}]{Me++05aa}
{Southworth}, J., {Maxted}, P.~F.~L., \& {Smalley}, B. 2005, A\&A, 429, 645

\bibitem[{{Southworth} et~al.(2007){Southworth}, {Wheatley}, \&
  {Sams}}]{Me++07mn}
{Southworth}, J., {Wheatley}, P.~J., \& {Sams}, G. 2007, MNRAS, 379, L11

\bibitem[{{Tinetti} et~al.(2010){Tinetti}, {Deroo}, {Swain}, {Griffith},
  {Vasisht}, {Brown}, {Burke}, \& {McCullough}}]{Tinetti+10apj}
{Tinetti}, G., et al. 2010, ApJ, 712, L139

\bibitem[{{Torres}(2013)}]{Torres13an}
{Torres}, G. 2013, Astronomische Nachrichten, 334, 4

\bibitem[{{Torres} et~al.(2010){Torres}, {Andersen}, \&
  {Gim{\'e}nez}}]{Torres++10aarv}
{Torres}, G., {Andersen}, J., \& {Gim{\'e}nez}, A. 2010, A\&ARv, 18, 67

\bibitem[{{Torres} et~al.(2012){Torres}, {Fischer}, {Sozzetti}, {Buchhave},
  {Winn}, {Holman}, \& {Carter}}]{Torres+12apj}
{Torres}, G., et al. 2012, ApJ, 757, 161

\bibitem[{{Torres} et~al.(2008){Torres}, {Winn}, \& {Holman}}]{Torres++08apj}
{Torres}, G., {Winn}, J.~N., \& {Holman}, M.~J. 2008, ApJ, 677, 1324

\bibitem[{{Tregloan-Reed} \& {Southworth}(2013)}]{TregloanSouthworth13mn}
{Tregloan-Reed}, J., \& {Southworth}, J. 2013, MNRAS, 431, 966

\bibitem[{{Tregloan-Reed} et~al.(2013){Tregloan-Reed}, {Southworth}, \&
  {Tappert}}]{Tregloan++13mn}
{Tregloan-Reed}, J., {Southworth}, J., \& {Tappert}, C. 2013, MNRAS, 428, 3671

\bibitem[{{Triaud}(2011)}]{Triaud11aa}
{Triaud}, A.~H.~M.~J. 2011, A\&A, 534, L6

\bibitem[{{Triaud} et~al.(2010){Triaud}, {Collier Cameron}, {Queloz},
  {Anderson}, {Gillon}, {Hebb}, {Hellier}, {Loeillet}, {Maxted}, {Mayor}, \&
  {Others}}]{Triaud+10aa}
{Triaud}, A.~H.~M.~J., et al. 2010, A\&A, 524, A25

\bibitem[{{Udalski} et~al.(2002){Udalski}, {Paczy{\'n}ski}, {Zebrun},
  {Szymanski}, {Kubiak}, {Soszynski}, {Szewczyk}, {Wyrzykowski}, \&
  {Pietrzynski}}]{Udalski+02aca}
{Udalski}, A., et al. 2002, AcA, 52, 1

\bibitem[{{Wakeford} et~al.(2013){Wakeford}, {Sing}, {Deming}, {Gibson},
  {Fortney}, {Burrows}, {Ballester}, {Nikolov}, {Aigrain}, {Henry}, {Knutson},
  {Lecavelier des Etangs}, {Pont}, {Showman}, {Vidal-Madjar}, \&
  {Zahnle}}]{Wakeford+13mn}
{Wakeford}, H.~R., et al. 2013, MNRAS, 435, 3481

\bibitem[{{Winn} et~al.(2010){Winn}, {Fabrycky}, {Albrecht}, \&
  {Johnson}}]{Winn+10apj3}
{Winn}, J.~N., {Fabrycky}, D., {Albrecht}, S., \& {Johnson}, J.~A. 2010, ApJ, 718, L145

\bibitem[{{Wright} \& {Gaudi}(2013)}]{WrightGaudi13book}
{Wright}, J.~T., \& {Gaudi}, B.~S. 2013, {in Planets, Stars and Stellar Systems.~Volume 3}, 489

\bibitem[{{Zahnle} et~al.(2009){Zahnle}, {Marley}, {Freedman}, {Lodders}, \&
  {Fortney}}]{Zahnle+09apj}
{Zahnle}, K., {Marley}, M.~S., {Freedman}, R.~S., {Lodders}, K., \& {Fortney},
  J.~J. 2009, ApJ, 701, L20

\end{thebibliography}

%%%%%%%%%%%%%%%%%%%%%%%%%%%%%%%%%%%%%%%%%%%%%%%%%%%%%%%%%%%%%%%%%%%%%%%%%%%%%%%%%%%%%%%%%%%%%%%%%%%%%%%%%%%%%%%%%%%%%%%%%%%%%%%%%%%%%%%%%%%%%%%%%%%%%
\end{document}